# Hypergraphs Demonstrate Anastomoses During Divergent Integration

**Bradly Alicea**[1,2]

**Abstract**

Complex networks can be used to analyze structures and systems in the embryo. Not only can we characterize growth and the emergence of form, but also differentiation. The process of differentiation from precursor cell populations to distinct functional tissues is of particular interest. These phenomena can be captured using a hypergraph consisting of nodes represented by cell type categories and arranged as a directed cyclic graph (lineage hypergraph) and a complex network (spatial hypergraph). The lineage hypergraph models the developmental process as an *n-ary* tree, which can model two or more descendent categories per division event. A lineage tree based on the mosaic development of the nematode *C. elegans* (*2-ary* tree), is used to capture this process. Each round of divisions produces a new set of categories that allow for exchange of cells between types. An example from single-cell morphogenesis based on the cyanobacterial species *Nostoc punctiforme* (multiple discontinuous 2-*ary* tree) is also used to demonstrate the flexibility of this method. This model allows for new structures to emerge (such as a connectome) while also demonstrating how precursor categories are maintained for purposes such as dedifferentiation or other forms of cell fate plasticity. To understand this process of divergent integration, we analyze the directed hypergraph and categorical models, in addition to considering the role of network fistulas (spaces that conjoin two functional modules) and spatial restriction.

# Introduction

During the development of an embryo, cells divide and differentiate, eventually forming tissues and physiological systems. This process has typically been characterized using tree-like structures, the most famous example being the lineage tree. The cells of an embryo can also be characterized as a complex network [1], and is useful in understanding phenomena such as intercellular signaling and geometric/spatial constraints. Furthermore, these embryo networks tend to diverge as cells differentiate, migrate, and form structures such as tissues and organs. The resulting functional and structural divergence is called *divergent integration* [2]. During divergent integration, components of the network diverge in both function and connectivity but retain a small number of connections while remaining part of the same system. These weak ties [3, 4] enable functional integration that defines complex organisms that exhibit modularity in the adult phenotype [5]. We will address the problem of establishing degrees of connectivity at multiple organizational and spatial scales. This will lead to developmental hypergraphs that approximate the biology of anastomoses (or categorical bifurcations) leading to and ultimately connecting different functional categories, modules, and tissues.

---

[1] OpenWorm Foundation, Boston, MA.  balicea@openworm.org
[2] Orthogonal Research and Education Lab, Champaign-Urbana, IL.

But how do we characterize this process across time and, more importantly, across cell births, deaths, and respecification of cell fate? A hypergraph structure will be used to graphically represent these processes, and provides a means to quantify these dynamics. The developmental hypergraph modeled here is a directed cyclic *n-ary* graph [6]. Each set of connections made between mother ($t$) and daughter ($t+1$) categories can capture both binary (two daughter categories) and proliferative (more than two daughter categories) modes of differentiation. This does not strictly mirror cell division in the lineage tree, but does contain that information. In general, all mother categories contain $N$ cells and daughter categories contain $n > 2N$ cells, regardless of the number of mother and daughter categories, respectively. Our hypergraphs exhibit partially independent systems [7], in which cells within a single functional category are organized in the same category. These categories can be refined after each cell division with limited interchange due to functional plasticity.

**Previous Work on Network Growth**

While our density bifurcation model is specific to capturing the spatiotemporal network dynamics of developmental cell lineages, earlier work in protein-protein interaction networks yielded the duplication-divergence model of network expansion [8]. Protein binding network in *Drosophila* [9] shows that networks can grow asymmetrically: *n*-cliques tend to duplicate and then diverge in terms of function. Growth of the network tends to involve the proliferation of relatively small components, and results in growth of the average network degree measure. More generally, the duplication-divergence model conforms to mean-field theory for various protein-protein interaction network topologies [10], yielding power-law behavior, scale-free properties, and hierarchical modularity, all key properties of real-world systems.

In developmental lineage trees, network expansion proceeds according to binary cell divisions, then binary categories of cell type. Over development, this results in a partially ordered network: cells diverge in terms of spatial location and function, but ultimately sort into categories that match the number of tissues in the adult form [11]. Cells and tissues are not ideally suited to a nodal identity due to the complexity and need to integrate this complexity into a functional identity. Thus, developmental processes produce a number of expanding hyperpaths [12] which coalesce into a hypergraph. This leads us to propose an alternative model of network expansion that is more closely allied with cellular development.

**Density Bifurcation Model**

The process of increasing connectivity in development proceeds by cell differentiation driving tissue differentiation, which in turn drives embryo network differentiation. The first stage involves cells dividing and migrating, resulting in increased connectivity. A greater number of cells results in a denser network topology. When the process of migration and differentiation are locally directed, such as a process where precursor cells act in a coordinated manner, it provides opportunities for this dense connectivity to be selectively broken apart. Thus as the function of cells diverges according to the rules of differentiation, selective migration and associated structural distinctions result in two interconnected networks. Additional cell migration resulting from this process enriches local network communities and cliques. The interconnected networks resulting from these two processes provide weak ties (functional interdependencies) between emerging tissues. Selected connections between what will become distinct categories are maintained due to the interdependence of tissues and organs in an organism, in addition to later differentiation of precursor cells at different points in time.

**Applying Hypergraphs to Embryogenesis**

Morphogenesis is broadly defined by two attributes: a reaction-diffusion process enabled by activation and inhibition [13], and differential inheritance of cell fate determinants [14]. Our model maps these constraints on developing cell populations to a hypergraph structure. Hypergraphs are graph structures where a group of vertices share the same edge (or hyperedge) with multiple other vertices simultaneously. To define hypervertices, we utilize what Makinen [15] calls the subset standard for hypergraph construction. Hypergraphs assembled using this method define the hypergraph structure as a collection of subsets originating from a superset of vertices. Hypergraphs also allow for multiple hyperedges to join a hypervertex containing multiple vertices [16]. To characterize cell lineage trees and other hereditary developmental relationships, directed hypergraphs [17] can be brought to bear. Directed hypergraphs are particularly good for modeling binary relations for subsets of a larger system [18], examples of which being cell division and differentiation.

Hypergraphs are also useful for connected entities that are connected due to shared context [19]. In development, shared context includes similarities such as genotype, phenotype, or functional state. While embryogenetic hypergraphs involve the characterization of multiple tissue types and anatomical structures, the application of hypergraphs to neuroscience is revealing. According to Jain [20], hypergraphs are ideal for modeling functional modules with different levels of granularity. Hypergraphs can also serve as a generalization of graphs such as connectomes, and define functional phenomenology in developmental neurobiology such as the emergence of tissues and organs [21]. Developmental hypergraphs can be spatially-distributed, unfolding in time, or a combination of the two.

To demonstrate hypergraphs in a developmental context we will present a hypothetical embryogenetic hypergraph (Figure 1) with the simplicity of *C. elegans* embryo [22] but with the regulative potential of vertebrate embryos [23]. Embryogenetic hypergraphs also represent the process of *anastomosis*. Anastomoses serve to connect two divergent structures. Examples in the adult human phenotype include valves between different chambers of the heart, or the sphincter and valve connecting the large and small intestine. In a network abstraction, anastomoses are cross-connections between subgraphs representing the differentiated subgraphs of an embryo network. These connections reveal exchanges of cells between subgraphs as they change identity from one functional cell category to another.

**Categorical Process of Divergence**

One way to characterize categorical transformation between subsets of vertices over time in the hypergraph is to utilize a category-theoretic model of embryogenesis [24]. In the case of embryogenesis, functional divergence results in a weakly tied together phenotype. Category theory allows us to build a hypergraph based on a formalism that stresses both cycles and transformations between categories at different points in time. Dorfler and Waller [25] propose the formalism $G = (v,e,f)$, where functions ($f$) are used to map between functionally-specific sets of vertices ($v$) and edges ($e$). Exchanges between categories are called *functors* [26]. While not featured explicitly in our model, operations between categories are called *monoids* [26]. Monoidal relationships are valuable in understanding transformative quantitative relationships such as cell doubling time (multiplicative) and processes such as differentiation, dedifferentiation, and respecification (additive, multiplicative, or fractional), where cells change from one fate to another as functional groups begin to diverge.

**Categorical Representation**

Connections between hypernodes can be represented using category theory, in particular a model of categories and functors (Figure 1). Each vertex (oval) in the hypergraphs shown in Figures 2 and 3 can also be considered a temporal-functional category. Divergence occurs over time as an exchange of cells from one category to another. Leaving aside the process of differentiation, which in regulative development can be spontaneous and independent of cell division, differentiation is defined by the transformation of a cell's category. Examples of multiple possible functors are shown in Figure 1. The functors *f(x)* and *f(y)* are particularly interesting in that they represent active differentiation (or respecification) at the same point in developmental time. This can occur due to differentiation after cell division or migration between subcategories occurring before the next round of cell divisions.

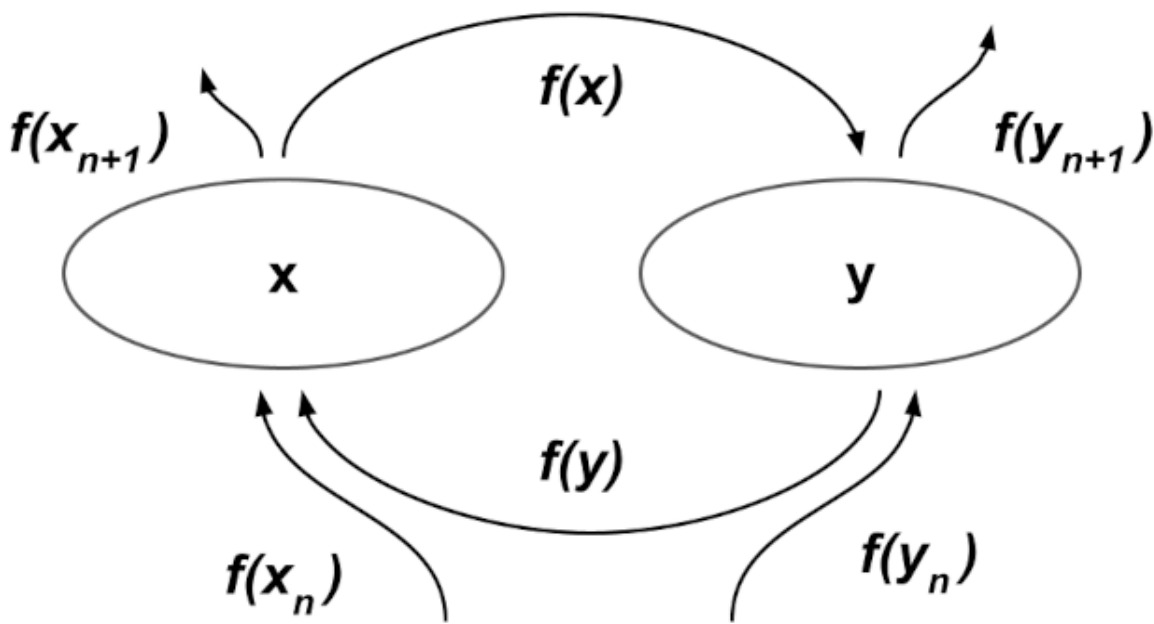

**Figure 1.** A category-theoretic representation of exchange between temporal-functional categories. Categories at current time are shown as *x* and *y*. Functors connecting these categories from previous and future cell division events are shown as $f(x_n)$, $f(y_n)$ and $f(x_{n+1})$, $f(y_{n+1})$, respectively. Cycles are represented by functors *f(x)* and *f(y)*.

**Source Data and Pseudo-data**

Multiple datasets and generated pseudo-data are used to illustrate the concepts introduced in this paper. Pseudo-data is used to generate the hypergraphs in Figures 2, 3, 4, 6, 7 and 8. Figures 2, 3 6, and 8 are based on the lineage tree and spatial relationships of a *C. elegans* embryo. Figure 4 is based on the growth rules and spatial organization of the cyanobacteria *Nostoc punctiforme*. Figure 7 is based on hypothetical distributions of transcriptional regulators and proteins for a 2-cell category (hypernode) output . Figures 5A and 5B are based on the following secondary data: the Figure 5A *C. elegans* data is taken from references [27, 28], while the Figure 5B *H. sapiens* data is taken from references [11, 29].

# Examples

**Hypergraph Analysis**

  Our hypergraph is represented as a directed cyclic graph, or a directed graph (or tree) that exhibits reticulating paths between subnetworks. In Figures 1 and 2, subnetworks are represented by the embryo and connectome. This can be contrasted with the germline, the branch path for which is isolated early from the rest of the network. As time increases, the number of nodes increases, and has parallels with increasing embryonic complexity.

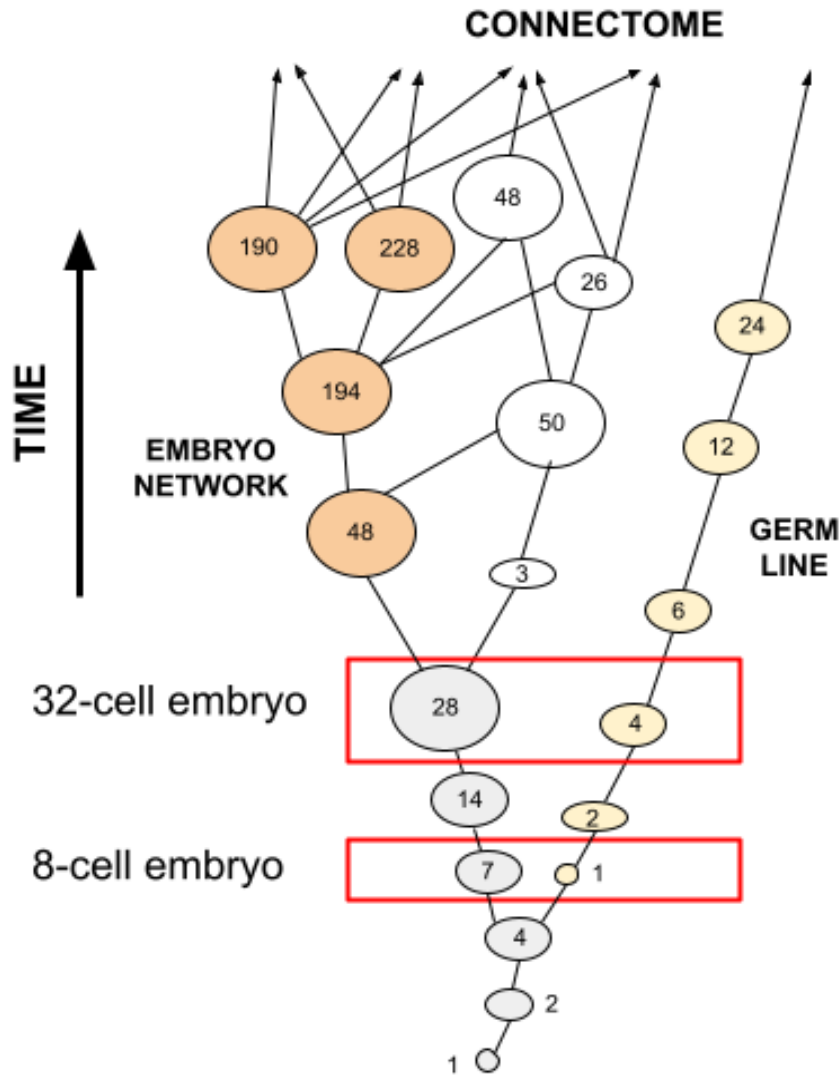

**Figure 2.** A hypergraph of an early-stage embryo similar to *C. elegans*. Yellow nodes: germ line, gray nodes: developmental cells, orange nodes: somatic cells (generic), White nodes: neural connectome.

Focusing on the numerous reticulation points shown in Figure 2, the connections between embryo subnetwork nodes and the connectome subnetwork nodes goes in a single direction (from precursor to differentiated cell types). While not shown here, paths can also return connectome cells to their original precursor population. This is useful for defining processes such as dedifferentiation [30] and other forms of cell plasticity. Such linkages can also reinforce functional ties between subnetworks.

**Standard lineage tree hypergraph**

In a lineage tree hypergraph representation, each hypergraph node features a proportion of cells in the embryo. As the network makes a binary ramification, the constituent cells in the mother node divide, their number doubles, and those cells are distributed at some proportion between the daughter nodes. For example, after the four cell stage in Figure 2, the network ramifies into two subgraphs: the embryo subnetwork (three cells) and the germ line subnetwork (one cell). While they do not always lead to differentiation, subsequent cell division events can lead to functional subpopulations that may be the source of subsequent differentiation events.

**Spatial hypergraph**

An alternate method for characterizing cell division and differentiation in development is by using a spatial hypergraph representation (Figure 3). In the spatial configuration, the categories are not arranged as they are in the lineage tree, but along the anterior-posterior (A-P) and left-right (L-R) axes. Upon cell doubling, the cells from each category in the *n* cell embryo are mapped via functors to cells in the *2n* cell embryo. This can be extended to the identification of differentiation trees and waves as shown in [31], Figures 22 and 23 .

One feature of a spatial hypergraph is that of *connected categories*. Connected categories [25] can be defined as an isomorphism of categories in the following way: every functor *f* that connects two morphisms $j_1$, $j_2$ results in equivalent non-empty categories $j_1$ and $j_2$. In the context of a developmental hypergraph, categories representing a common tissue type (e.g. muscle or germline) can be considered isomorphic, each category consisting of spatially local cells with similar attributes. Cellular attributes can range from identity to molecular complements (common genetic regulatory outputs). Connected categories capture functional commonalities that transcend spatial localization. For example, all categories of neural tissue are considered to be connected, along with categories that are ancestral to or descended from these spatially connected categories. Connected categories of muscle can be descended from undifferentiated developmental cells, or serve as the ancestors of a smaller set of connected categories. In this case, the lineage and developmental hypergraphs intersect with functors linking the isomorphic and non-isomorphic categories across space and time.

**Developmental Hypergraph for Single-cell Morphogenesis**

Instances of single-cell morphogenesis can be found in organisms as wide ranging as Diatoms and Cyanobacteria. We can employ the developmental hypergraph method in much the same way as for the examples shown in Figures 2 and 3. This type of morphogenesis is extraordinarily diverse and provides unique challenges to our model. Diatom morphogenesis is defined by periodic molecular self-assembly of silicates [32], while cyanobacteria morphogenesis is marked by reversible differentiation resulting in a chain with differentiated cell types forming a nonrandom pattern [33, 34]. In both of these cases, development proceeds not according to a programmed procession of tissue differentiation, but through highly local interactions determined by molecular interactions (not cell-dependent).

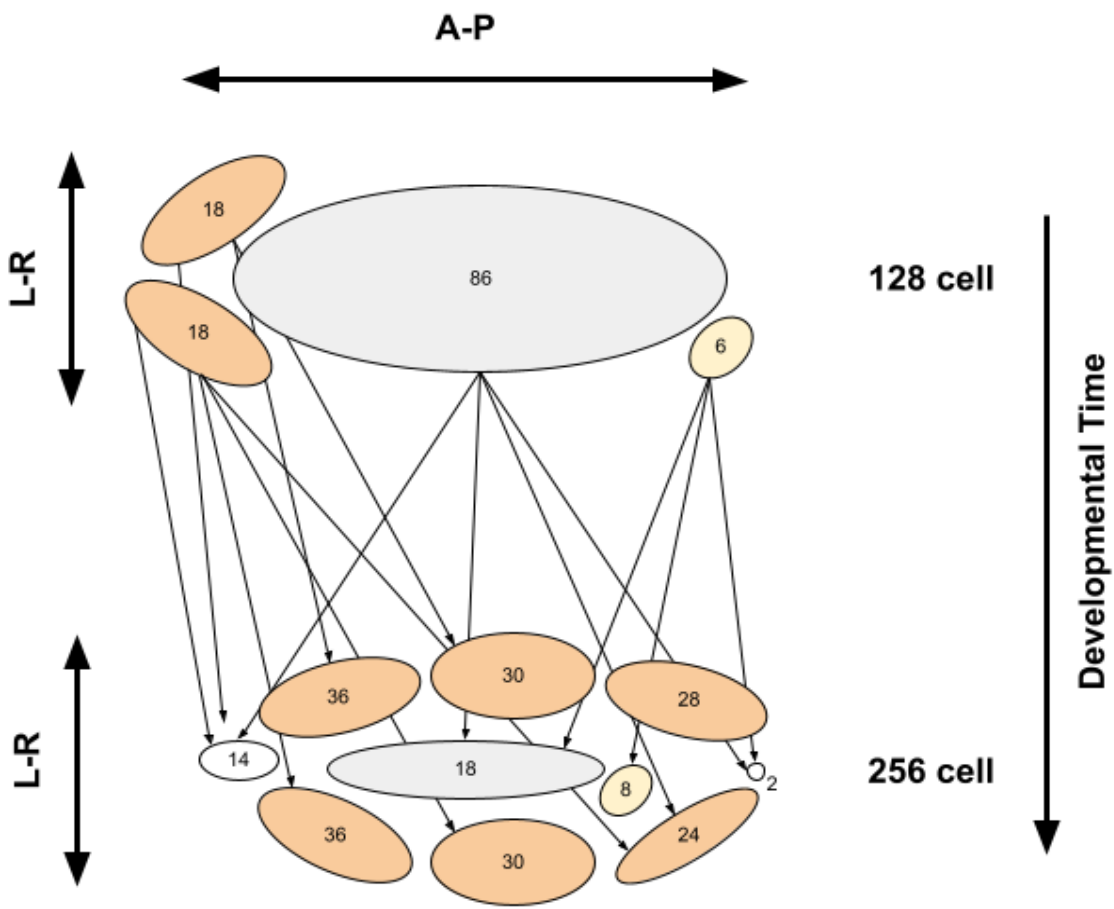

**Figure 3.** An alternate version of the developmental hypergraph focusing on spatial differentiation. A-P: anterior-posterior, L-R: left-right. Yellow nodes: germ line, gray

nodes: developmental cells, orange nodes: somatic cells (generic), White nodes: neural connectome.

Figure 4 demonstrates the configuration for a cyanobacterial colony based on *Nostoc punctiforme*, the growth and form of which forms an spatially extended chain of heterocysts [35, 36]. At certain locations in the chain, there exist sites that expand, forming a network fistula between the two other parts of the chain. In the hypergraph model, our directed hypergraph does not have an endpoint, as growth can continue as long as the cells are actively dividing. The only way we terminate our hypergraph is through the death of all cells in the graph. Cell division leads to differential inheritance of factors shaping restricted cellular differentiation [35]. It follows then that introducing molecular perturbations in descendent cells can lead to a disruption of pattern formation. Mutations in the transcriptional regulator *patN* can delay and randomize the nonrandom pattern of the heterocyst chain, with *patS* and *hetN* acting more generally to inhibit cell differentiation [37]. At least 12 proteins serve to control development and patterning in all cell types in this system [36], which can serve to define our spectral output for each hypergraph category.

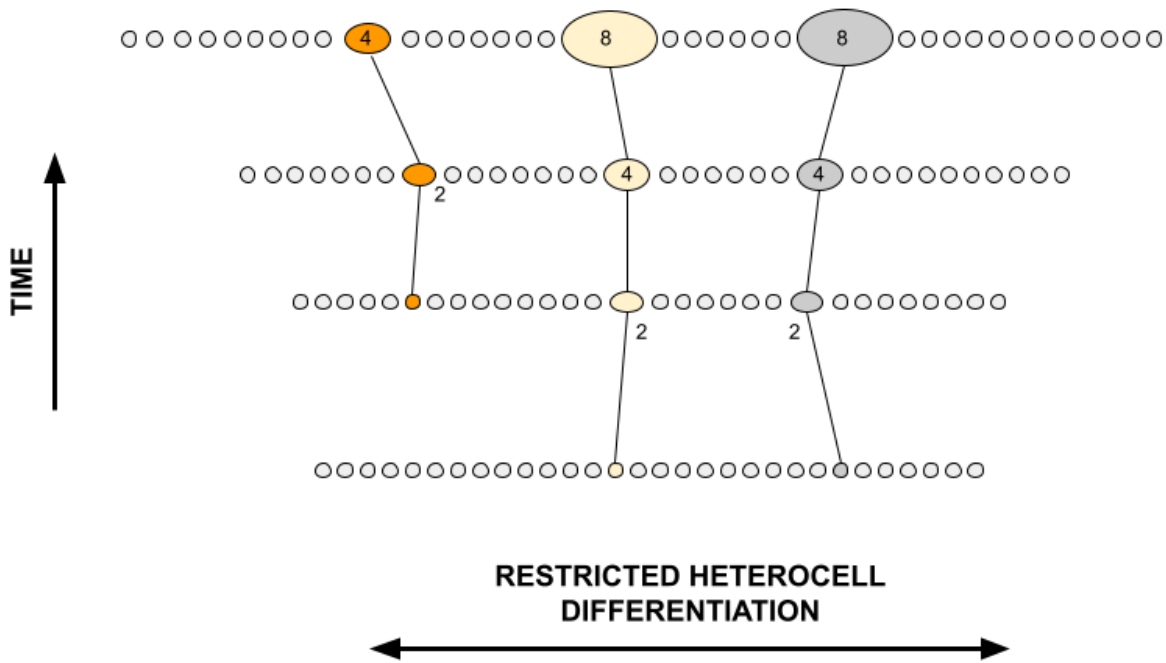

**Figure 4.** A developmental hypergraph representing restricted single-cell morphogenesis modeled as a cyanobacterial chain common in the species *Nostoc punctiforme*. In this depiction, a single cell expands into a category of two cells, four cells, and eight cells. Each horizontal array of cells represents a potential doubling

event. The yellow and gray lineages begin on the bottom array, while the orange lineage begins after a single doubling event has expanded the length of the chain.

**Functional Anastomosis as Overlapping Profiles**

To investigate the subtleties of cell fate identity, transcriptional profiling can be used to understand temporal transitions between phenotypic states or subcategories that refine function [38, 39]. Transcriptional switches enable potential anastomoses by pushing the phenotype from one state to another. In a network with multiple threshold criteria, cells can belong to multiple categories simultaneously. Furthermore, common sets of transcriptional changes among multiple cells, perhaps resulting from intercellular signaling, can define a category-theoretic functor. These various types of overlapping profiles can provide some of the basis for weak ties in embryo networks, particularly when hypergraph subcategories are also spatially interdigitated.

**Spectral Output of Cell Type Categories**

At the end of development (our adult phenotype), the hypergraph representation yields a distribution of cell types and frequency of cells in those categories. This results in a frequency spectrum that is shown in Figure 5. We can estimate what a frequency spectrum should look like from data from the adult hermaphrodite *C. elegans* [27, 28] and the adult male and female *H. sapiens* [11, 29]. In *C. elegans* (FIgure 4A), the mean volume is calculated (volume per nucleus count), while in *H. sapiens* (Figure 5B) the mean size is calculated (mass per cell count normalized by 10 + log).

We further expect three types of idealized relationships for frequency spectra. The first is where all cell/tissue types are equally represented. We might observe this as cases of single-cell morphogenesis among organisms such as diatoms or cyanobacteria, where the output spectrum for a lineage is the distribution of transcriptional regulators and proteins, not cell properties (see Figure 5). In the equiprobable case, cell types are well-mixed and the phenotype is not specialized. The second case is hierarchical dominant signatures, which is partially observed in Figure 5. Some cell types are much more common than others, and implies a set of functional relationships. *C. elegans* has many more intestinal cells than any other type, while in *H. sapiens*, PNS (peripheral nervous system) and adipose cells are much more common than skin cells. In the third case, multimodality occurs when there are different and perhaps separable functional regimes. These functional regimes could be due to symmetry-breaking (asymmetries) during embryogenesis.

In Figure 6, a spectral output for one category of developmental cells shown in Figure 1 are analyzed using pseudo-data. This category contains 194 cells, and contains cells that exhibit Anterior-Posterior (A-P) polarity as well as cells that are primed for differentiation in the next round of division. This category is slightly biased towards anterior cells (versus posterior cells), while also containing a smaller fraction of differentiation-primed cells, ready to become connectome cells in the next round of division. For the single-cell hypergraph of cyanobacteria morphogenesis (Figure 3), we can look at the spectrum from a single category using pseudo-data. In Figure 7, we consider the output of a two-cell category (2-*nary* tree with a single level) to consider the asymmetry of molecular factors in each constituent cell. Our hypothetical spectrum includes a complement of 16 factors. Of note is the asymmetry between cells in terms of transcriptional regulator and protein count. In Figure 7, *Bm, Sr, and Tc, and Vr* show the greatest disparity between cells in the category. This reflects the differential inheritance of factors in the cytoplasm of one cell over another after cell division.

## Discussion

Hypergraph representations of developmental processes, particularly embryogenetic processes, allows us to model the embryo at both the cellular and extracellular/tissue/organ level. This model of hypergraph exchange and subgraph partitioning can be further understood with the density-bifurcation model, which proposes a developmentally specific model of node attachment. This work is particularly relevant to building spatiotemporal representations of embryo dynamics.

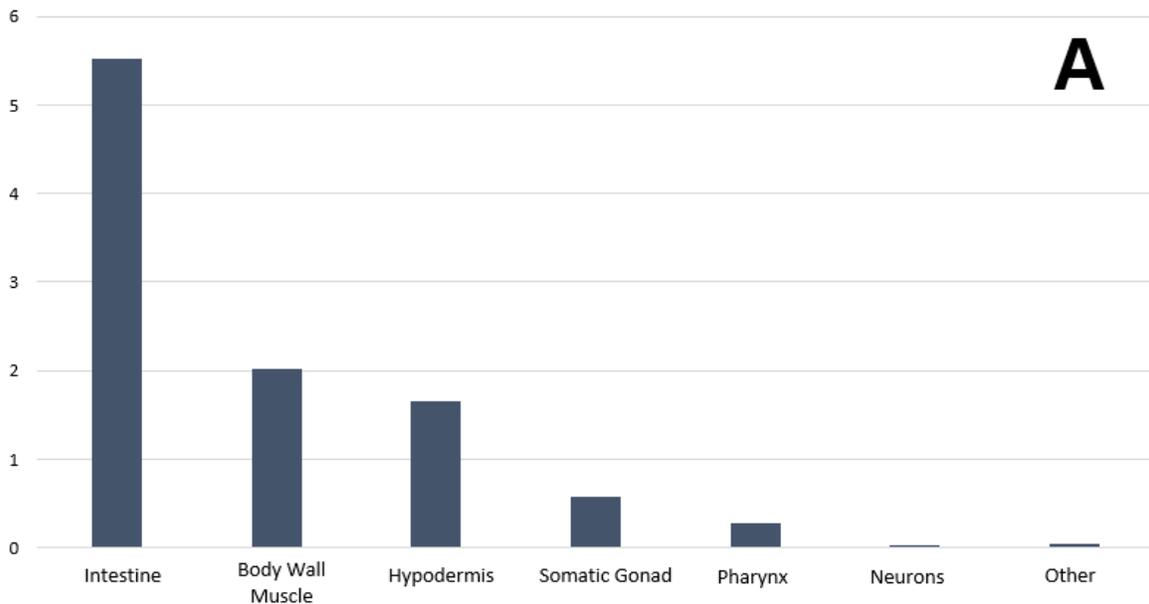

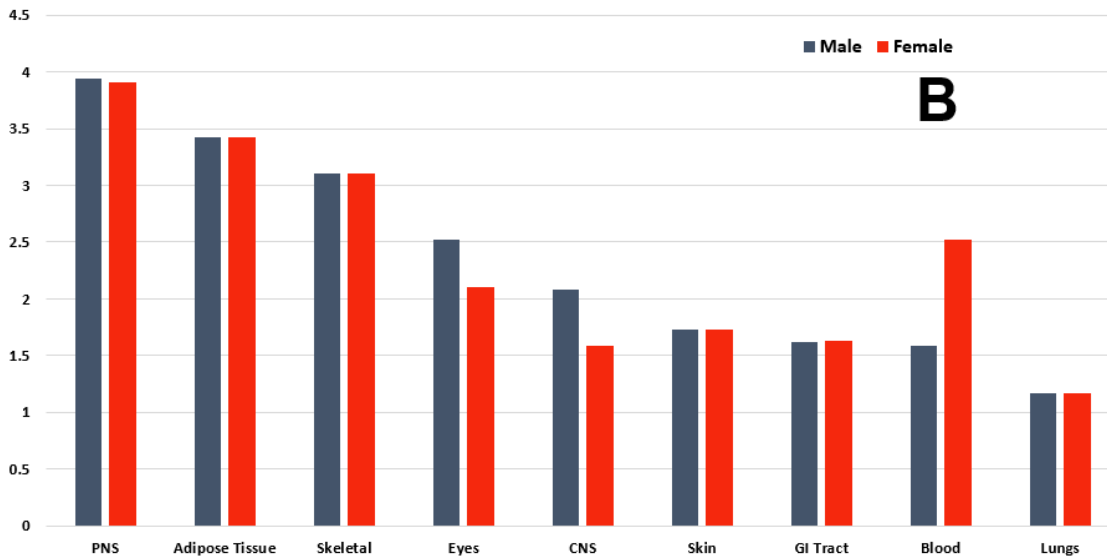

**Figure 5.** Spectral output for different cell type categories. Panel A features different tissue types in the *C. elegans* hermaprhodite. Panel B features different male and female tissue types in *H. sapiens*. Data sources described in Source Data and Pseudo-data subsection of the Introduction.

Our hypergraph and categorical models are not explicitly rooted in the spatial identity of cells. However, this method is ideal to represent the diversity of biological cell types [40]. Nevertheless, a spatial version of the hypergraph can be derived that demonstrates changes in spatial and cell type differentiation over time. The primary concern of the analysis shown here was to understand cell sorting and functional divergence over time. However, functional divergence also occurs with precise spatial restriction and has consequences for pattern formation [41]. While the lineage-based hypergraph shown in Figure 1 is ideal for identifying the emergence of substructures in the embryo (such as the connectome), the spatial-based hypergraph may be good for visualizing the phenotypic outcomes of spatial transcriptomics [42] or coordinated signals such as differentiation waves [31].

**Network fistulas and developmental topology**

Fistulas are anastomoses that form abnormally between two parts of the phenotype. Sometimes this is due to a developmental mutation [43], but sometimes network fistulas and associated cell state transitions can serve as the grist for evolutionary change [44]. In network topologies, fistulas may emerge as a consequence of rewiring of embryo networks for functional specificity. The developing *C. elegans* connectome features several cavities [45] that might point to how spatially explicit conduits between functional subdivisions might be maintained. More generally, spatial processes in the network can be better summarized by topological methods [46]. In single-cell morphogenesis, the expansion of a chain via a single heterocyst can create a fistula between two non-expanding regions of the chain.

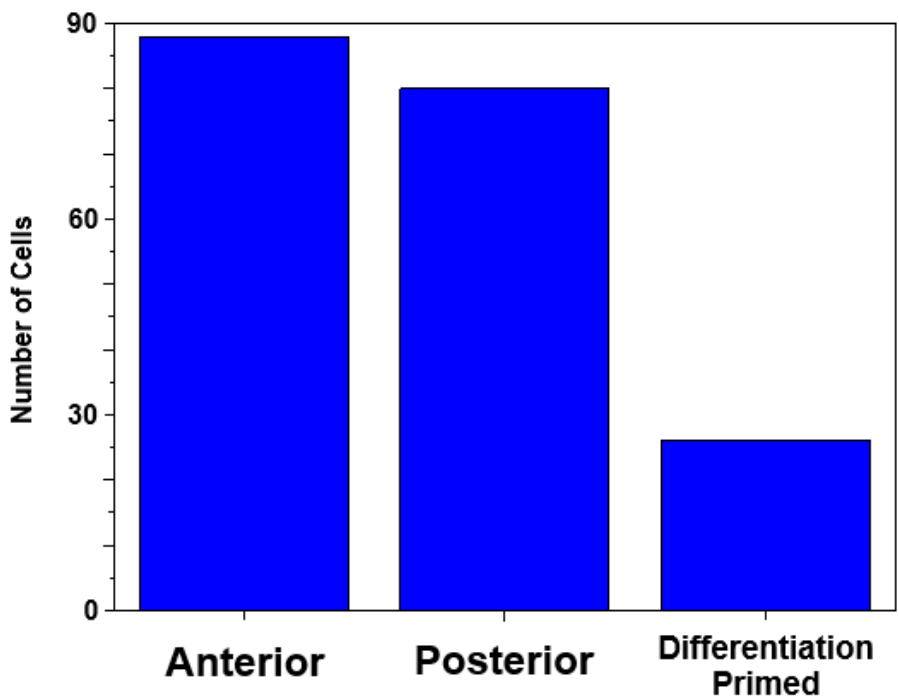

**Figure 6.** Hypothetical spectrum (modeled using pseudo-data) of cell subcategories in category of size 194 in the developmental cell 2-*ary* graph.

**Underlying processes of division and differentiation**

While differentiation between different cell type categories is ultimately under transcriptional control, the timing of cell division is cell cycle dependent. While this is also partially regulated by transcriptional activation [47, 48], a computational model can model this process as the relative length of time it takes to move between different cell cycle checkpoints [49]. In treating different components of the cell cycle as a variable timing mechanism, a greater diversity of a given developmental process can be more faithfully captured. Cell cycle models can be used to define individual precursor subsets in the hypergraph, or the constituent vertices (cells) that are exchanged between subsets via functor. Other biological functions might also be used to define or refine the components of the hypergraph model.

**Future Directions**

One advantage of this method is its amenability to the incorporation of new data types. Single cell transcriptomics [50, 51] can be used to provide sharper definitions to our categories and anastomoses between categories using more precise functors. On the other hand, different data types can often produce signals that point us in different directions. This is why we combine hypergraphs with categorical techniques: the branching and ramification structure of our directed hypergraph allows for intermediate states to be represented and analyzed using the spectral output. Our categorical representation allows for us to classify different spectral outputs. We predict three main types of output based on size-abundance and other functional tradeoffs, which is the product of developmental history enabled by our expanding hyperpaths.

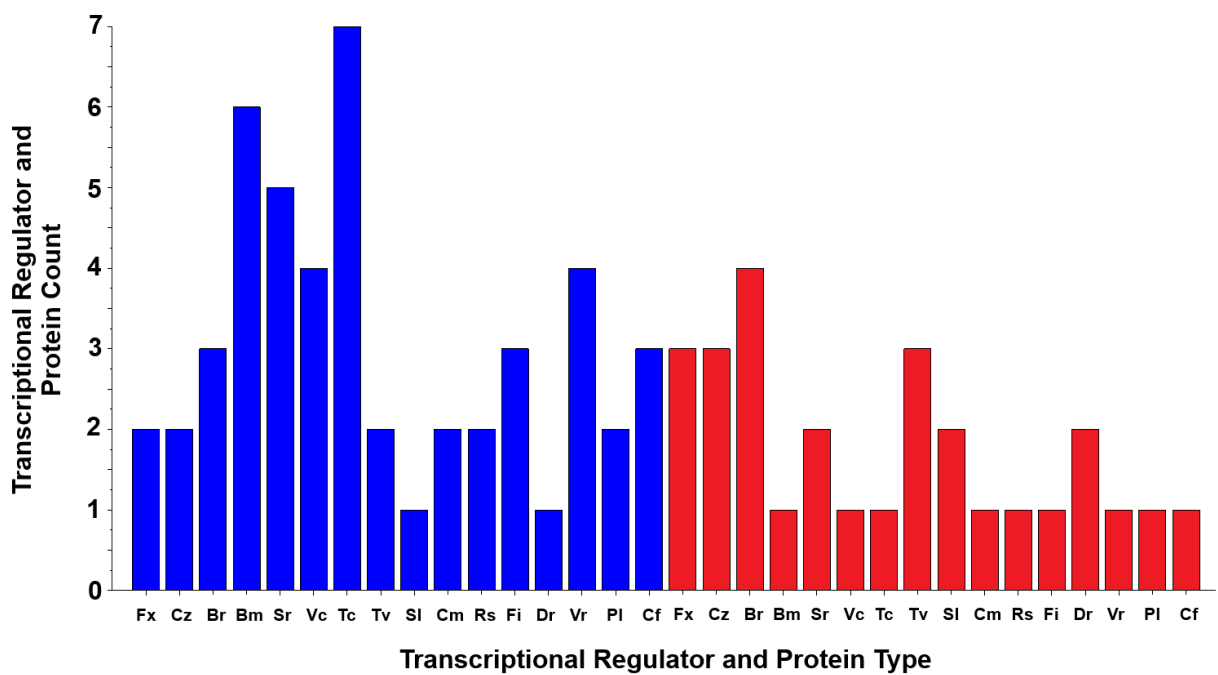

**Figure 7.** Pseudo distributions that define the 2-cell category (hypernode) output of transcriptional regulators and proteins. Identities are hypothetical, but defined in each cell in order: *Fx, Cz, Br, Bm, Sr, Vc, Tc, Tv, Sl, Cm, Rs, Fi, Dr, Vr, Pl, and Cf*. The spectrum shown here is for a category of size 2. Blue = Transcriptional Regulators and Protein spectrum from Cell 1, Red = Transcriptional Regulators and Protein spectrum from Cell 2. Types (16) are distributed differently in the two cells.

We can also use our proposed hypergraph model to understand aspects of development outside of embryogenesis. For example, we can view developing connectomes as a directed, embodied hypergraph by connecting the structural description shown in Figure 2 to a sensory input and a motor output. Each intermediate node in-between the input and output can be characterized as a hypernode that can propagate as a lineage with corresponding increases in the number of cells contained within that particular category (Figure 8).

**Broader Developmental Processes and Emergent Processes**

The directed, embodied hypergraph representation captures a number of broader developmental processes, including but not limited to: growth heterochrony, sequence heterochrony, and critical periods. Growth heterochrony involves the rate of growth during development. Growth rate can change due to systematic delays in the onset of growth, or early termination of growth. In an organ (analogous to our hypernodes), size can be modified by limiting the proliferation of cells during development [52]. In a similar fashion, sequence heterochrony involves changes in the order of events that unfold during development [53]. Changes in developmental order can lead to radically altered phenotypes, including network topologies. Finally, critical periods [54] provide a window of enhanced plasticity during which significant biological changes or informational acquisition can occur. In the case of our hypernodes, changes in the number of daughter categories, or the number of constituent cells in descendent hypernodes can be affected depending on when expansion of the network occurs.

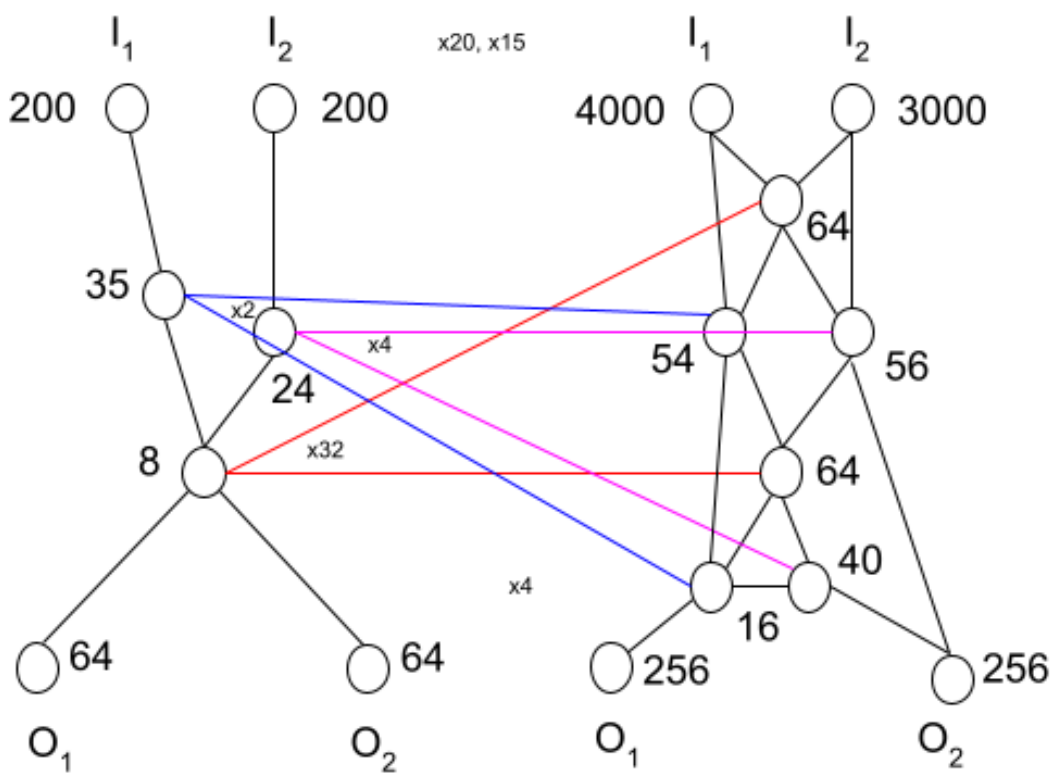

**Figure 8.** Two directed, embodied developmental hypergraphs for the same organism at adjacent periods in time. Set In features each sensory organ, a single hypernode that can contain many individual cells. Set On features each motor effector, which is a hypernode also including many cells. All other hypernodes represent interneurons at a particular location. The colored lines represent differentiation events: the hypergraph on the left features the mother hypernode, while the hypergraph on the right features

daughter hypernodes containing daughter cells of different categories. Colors are arbitrary. Large numbers are the number of cells in a hypernode, small numbers ($xn$) represent the magnitude of change between the number of cells in the ancestral and descendent hypernode, respectively. This can reveal both the speed of development and asymmetries between bilateral sensory and motor organ hypernodes (e.g. change in $I_1$ and $I_2$).

Aside from developmental processes, our directed hypergraphs allow us to capture emergent processes, as well as collective decisions made by cell populations. In the case of an emergent process, a directed hypergraph allows us to account for the additive components of an emergent structure, thus pointing us towards the unspecified superadditive components. Collective decisions can also be accounted for by using a decision-making kernel for every functor, which acts to determine the likelihood of each transformation from one category to the other. We can use these types of directed hypergraphs to model non-biological systems as well. One interesting application is developing an architecture for bioinspired neural networks [55]. Neural networks, particularly graph neural networks (GNNs), might benefit from the directed and modular properties of the developmental hypergraph representation.

# Acknowledgements

Thanks and gratitude go to Dr. Richard Gordon for his insights on differentiation waves and the DevoWorm group for their comments and suggestions more generally. Additional acknowledgement to the conferences that have retained a virtual/hybrid component, as this format has helped immensely in the development of our organization.